# Unusual Dielectric Relaxation in Lightly Doped n-Type Rhombohedral $BaTi_{0.85}Zr_{0.15}O_3$:Ta Ferroelectric Ceramics

*Jun Xu[*] and Mitsuru Itoh*

Materials and Structures Laboratory, Tokyo Institute of Technology, 4259 Nagatsuta, Midori, Yokohama 226-8503, Japan

**Abstract**

The dielectric properties of tantalum doped $BaTi_{0.85}Zr_{0.15}O_3$ were investigated. The conventional solid state reaction method was used to make the ceramic samples. The composition was close to the pinch point of the three structure phase transitions in $BaTiO_3$. Powder X-ray diffraction result indicated that the samples were in very slightly distorted rhombohedral structure at room temperature, and the dielectric measurement showed only cubic-to-rhombohedral phase transition occurred. Instead of a sharp peak characteristic of a ferroelectric (FE) transition, the large dielectric response was similar to those in nonferroelectric perovskites $CaCu_3Ti_4O_{12}$, $AFe_{1/2}B_{1/2}O_3$ (A=Ba, Sr, Ca; B=Nb, Ta, Sb) and IBLC (Internal Barrier Layer Capacitance) $BaTiO_3$ based FE ceramics. However, unlike its FE analogue, the conventional $BaTiO_3$ based IBLC ceramics, our samples showed weaker temperature dependence of relative permittivity which spans over a wide temperature range. A Maxwell-Wagner type relaxation in the FE state was observed. Moreover, another relaxation mode appears from the vicinity of FE phase transition temperature $T_C$, and extends over a narrow temperature range, with unusual slowing down of relaxation rate as temperature rises, which is contrary to what is commonly observed. We associate these unusual phenomena with the very slightly distorted rhombohedral structure of our samples and the presence of the grain boundary barrier layer where oxygen vacancies distribute non-homogeneously.

[*] Corresponding author. Present address: junxu@hiroshima-u.ac.jp

## Introduction

Mixed perovskite oxides exhibit a rich variety of dielectric relaxation phenomena at the low and intermediate frequency ranges (≤ 1 MHz). This behavior of the frequency dispersion of dielectric permittivity can occur at very low and very high temperatures, depending on the physical mechanism and the characteristic thermal activation energy. The origins can be both intrinsic (determined mainly by the lattice symmetry) and extrinsic (depending on the microstructures which can be modified by sample processing procedures).

In the well established off-center ion model, relaxation occurs through the hopping of the off-center ions among several equivalent energy minima which are determined by parameters such as the crystal symmetry, external field and the local environment. This type of dipole relaxation is responsible for most of the observed dipole dispersions, such as, in doped quantum ferroelectrics $KTaO_3$, $SrTiO_3$, etc., as discussed in reference 1.

The hopping of charge carriers (electrons and polarons), is also shown to make much contribution to dielectric response other than conductivity,[2-3] and may interact with the relaxing species in the lattice, resulting in diffuse dielectric anomalies.[2] The charge carriers may be from dopant ions and/or oxygen vacancies.

Oxygen vacancy is often unavoidable in perovskite oxides, especially under a high firing temperature and/or low partial oxygen pressure. The contribution of oxygen vacancies to dielectric relaxation can be 3-fold. The first is that oxygen vacancy can hop in the lattice, leading to the associated dipolar moment reorientation, which is in the frame of off-center ion model.[4] The second is by contributing conducting electrons through ionization. The conducting electrons (or small polarons in some cases), hop from one localized state to another in the lattice, equivalent to the reorientation of effective electric dipoles.[3,5] The third is that it can also manifest itself through one type of interfacial polarization mechanism, the dielectric relaxation of which is Maxwell-Wagner (MW) relaxation.[6] This is realized by different level of oxygen deficiency in different regions within the sample, leading to electric heterogeneity in a crystallographical single phase. The grain boundary in polycrystalline ceramics and various defect structures (such as twin boundary, dislocation) in single crystals play an important role in this mechanism.

The MW effect usually leads to dielectric relaxation with very high dielectric permittivity, and was shown to be the main mechanism in the recently found perovskites $CaCu_3Ti_4O_{12}$ [7-9] (CCTO) and $AFe_{1/2}B_{1/2}O_3$ (A = Ba, Sr, Ca; B = Nb, Ta, Sb)[10] (AFBO). Both two are non-ferroelectric and exhibit weak temperature dependent high dielectric permittivity over a wide temperature range which subsequently relax out at low temperatures. The very high dielectric permittivity arises from the same mechanism as in the internal barrier layer capacitor (IBLC) ferroelectric (FE) ceramics, in which the energy barrier at the grain boundary is important. IBLC mechanism has been utilized commercially in mostly $BaTiO_3$-based ceramics.[11] However, there, the high permittivity in the vicinity of FE phase transition temperature $T_C$ is more governed by FE response and varies more strongly with temperature, comparing to nonferroelectric CCTO and AFBO.

In all the dielectric relaxations described above, the relaxation rate increases as temperature increases, as can be expected from a thermally activated process. We report here an unusual relaxation mode in which the relaxation rate decreases as temperature increases. It was observed in the lightly donor doped rhombohedral FE $Ba(Ti_{0.85}Zr_{0.15})O_3$:Ta ceramics sintered in air, in a narrow temperature range starting from the vicinity of $T_C$. And the dielectric response of the samples show similar behavior as in CCTO and AFBO, namely, weak temperature dependent high relative permittivity over a wide temperature range which relax out at low temperature, which is a second interesting observation: MW relaxation in the FE state. We ascribe these unusual phenomena to the very slightly distorted rhombohedral crystal structure of our samples and the presence of grain boundary barrier layers where oxygen vacancies distribute non-homogeneously.

## Experimental

The conventional solid state reaction method was used to make the Ta doped $Ba(Ti_{0.85}Zr_{0.15})O_3$ samples. The nominal composition is $Ba_{1-0.5x}(Ti_{0.85}Zr_{0.15})_{1-x}Ta_xO_3$ with x=0, 0.08%, 0.2% and 0.5%. Stoichiometric $BaCO_3$ (99.99%), $TiO_2$ (99.9%), $ZrO_2$ (99.99%), $Ta_2O_5$ (99.9%) were thoroughly mixed in alcohol, calcined in 1250°C–1300°C with intermediate grindings. Pressed samples in disk shape with 10mm diameter were sintered at 1400°C or 1500°C for 6 hours in air after holding at 500°C for 8 hours to burn out the binder. Finally, the samples were furnace cooled to room temperature.

The phase purity was identified by X-ray diffraction using Cu Kα radiation at 50kV and 300mA with a rotating anode diffractometer M18XHF from MAC Science. Gold electrodes were fired on the polished surfaces of the samples for dielectric measurements. The dielectric permittivity was measured in the temperature range of 5K-400K using a HP4284A LCR meter. To avoid non-linear effect, small probing ac field ~50v/m was used. Resistivity measurement in the same temperature range was carried out using a Physical Property Measurement System from Quantum Design by a standard 4 probe method.

## Results and Discussion

The pinch effect for the three phase transitions in $BaTiO_3$ upon Zr substitution for Ti is well known. The three $T_C$'s tend to pinch together at certain amount of Zr substitution above which only cubic-rhombohedral phase transition occurs. The composition of our samples is just close to this pinch point in the $Ba(Ti,Zr)O_3$ phase diagram.[12] Careful X-ray diffraction measurement at room temperature show our samples are pseudo-cubic with very small rhombohedral distortion. Figure 1 is a result for the 0.08% Ta doped sample (denoted as BTZ:Ta008 hereafter.), which was taken in a step scan mode with step size 0.02° and counting time 2 seconds. We note that, the split of <hkl> reflections is so light that it is not discernible at our routine continuous scan mode for phase identification. The very light split, or more exactly, the appearance of a shoulder, for high angle reflections, such as the characteristic <222> reflection at around 2θ ~ 82.7°, can only be observed at much slower scan speed or step scan mode like (0.02°, 2sec). Apart from the pseudo-cubic peaks, no discernible unindexed peak was observed. Accordingly, dielectric permittivity measurements show only one sharp peak characteristic of FE transition at 343 K for $Ba(Ti_{0.85}Zr_{0.15})O_3$ sample, confirming that only one cubic-rhombohedral phase transition exists with $T_C$ = 343 K for x=0.

For the lightly Ta doped samples (x=0.08%, 0.2%, 0.5%), however, the temperature dependent dielectric permittivity (ε'(T)) behaves drastically different. Instead of a sharp peak indicative of the FE transition, ε'(T) shows weak temperature dependent permittivity at a high value (~$10^4$) over a wide temperature range from above $T_C$ (see below) to ~100K, below which a large drop of two orders of magnitude occurs with large frequency dispersion. A typical result is shown in Figure 2 for BTZ:Ta008 . The behavior is very similar to that in CCTO and AFBO, However, compared to those two cases, in our samples, ε' varies more with temperature and frequency. In addition, at high frequency, (e.g. 1 MHz), a small peak anomaly at ~340K can be clearly seen, which is the manifestation of the cubic-rhombohedral FE phase transition (see below).

The imaginary part of permittivity ε'' shows two peaks with frequency dispersion, indicating two relaxation modes (Figure 2(b)). The low T mode corresponds to the 100-fold sharp drop of ε'(T) step, and the peak shifts to higher temperature with increasing frequency. The high T mode begins to appear around $T_C$. Surprisingly, its peak shifts to higher temperature with decreasing frequency, which is contrary to commonly observed dielectric relaxations.

Dielectric relaxation can be more clearly seen from frequency dependent behavior. Figure 3 shows the real part of permittivity as a function of frequency at fixed temperatures. ε'(f) generally shows two plateaus at low and high frequency ranges respectively. The drop from the high plateau to the low one occurs at increasingly lower frequency when the temperature decreases, exhibiting a large frequency dispersion region. In addition, there is a minor intermediate plateau appearing in the ε'(f) curves for temperatures above 330 K, a manifestation of the weak high T mode relaxation process.

Accordingly, the frequency dependent imaginary permittivity ε''(f) shows two distinct behavior at two different temperature regions (Figure 4). At T<330K, ε''(f) curves show only one distinct peak which shifts to higher frequency with increasing temperature, corresponding to the low T mode in Figure 2. At T>330K, a second, weaker peak gradually evolves from the high frequency side and migrate to low frequency side with decreasing peak amplitude as temperature increases, which corresponds to the high T mode in Figure 2. Thus, the results show that there is only one relaxation process (low T mode) below 330 K, and two (low T mode + high T mode) above 330 K. The two modes above 330 K can be distinctively decomposed after resolving each set of peaks by curve fitting with two peaks. The relaxation frequency for each mode can thus be clearly extracted, which are plotted in Figure 5.

The low T mode was found to be a Debye-type relaxation process from Cole-Cole plot (not shown here). And the Arrhenius plot in Figure 5 shows that the relaxation rate ω (=2πf) at low temperatures follows the Arrhenius law $\omega=\omega_0\exp(-\Delta_d/kT)$, with the activation energy $\Delta_d$=0.128ev, and the pre-exponential frequency $\omega_0=1.9\text{x}10^9$Hz. The latter value indicates that the relaxation motion of low T mode is not an ionic process, and is comparable to that reported in CCTO.[9] Above 200K, the relaxation rate of low T mode deviates from the low temperature Arrhenius fit, which is found to be linked to the overall conductivity behavior (see below). On the other hand, contrary to the low T mode relaxation which spans over a wide temperature range, the high T mode relaxation occurs in a narrow temperature range starting from the vicinity of $T_C$ and exhibits unusual slowing down of the relaxation rate with temperature rising.

We found that the observed dielectric response in lightly doped samples strongly depends on sintering and post-sintering annealing conditions. The color of the sample without Ta doping is light yellow, whereas dark blue-gray for the lightly doped ones indicating a reduced state. If sintering the sample in a small flow of oxygen instead of in air, then we got light yellowish samples with only one sharp ε' peak characteristic of FE phase transition, and there is no dielectric relaxation observed. The FE phase transition temperature $T_C$ shows systematic decrease with increasing dopant concentration, indicating the incorporation of Ta dopant into the lattice. The $T_C$ for each lightly doped sample sintered in oxygen flow is coincident with the peak anomaly at high frequencies for the air-sintered samples (see Figure 2), indicating the peak anomaly at high frequencies is due to the FE transition. Annealing the air-sintered samples in a small oxygen flow also tends to diminish the large ε'(T) plateau and the associated frequency dispersion, with increasing FE phase transition feature at high frequencies. The annealing effect is more prominent for higher doped samples, e.g. the 0.5% Ta doped one, but little for the most lightly doped, e.g. BTZ:Ta008. The details will be published elsewhere.

The crucial dependence of the dielectric properties in sintering and post-sintering annealing condition as described above strongly suggests that the observed phenomena, namely, the enhanced large relative permittivity in a wide temperature range with weak temperature dependence, Debye-type relaxation at low temperatures in the FE state, and the unusual slowing down of the relaxation rate with increasing temperature for the high T mode which occurs in a narrow temperature range starting from the vicinity of $T_C$, stem from ceramic microstructure which is related to oxygen deficiency. As in CCTO[7-8] and AFBO[10], and IBLC $BaTiO_3$ based FE ceramics[11], the observed dielectric response is caused by the formation of boundary barrier layers due to a microstructure consisting of semiconducting grains and insulating grain boundaries. It's known that, semiconducting grains are produced by losing oxygen during sintering at high temperature, and the subsequent cooling process after sintering results in reoxidation along grain boundaries, yielding the insulating nature of grain boundaries. The absorbed oxygen at the grain boundaries are electron traps, or acceptors, which attract conducting electrons from the n-type semiconductive grains, creating a depletion layer with built in energy barrier, in the vicinity of grain boundaries. In $BaTiO_3$ based FE ceramics, the presence of grain boundary barrier layers results in the well known IBLC effect and PTCR (Positive Temperature Coefficient of Resistivity) effect,[13-14] which, we believe, is also a key factor underlying the observed unusual dielectric response in our samples.

From semiconductor theory, the width of the depletion layer at the grain boundary is:

$$b = n_s/2n_0 \tag{1}$$

where $n_s$ is the occupied grain boundary acceptor states, and $n_0$ the free charge carrier density. The height of the potential barrier $\phi$ which is built in with the depletion layer, is expressed as:

$$\phi = en_s^2/(8\varepsilon_r\varepsilon_0 n_0), \tag{2}$$

where e is the electron charge, $\varepsilon_r$ the relative permittivity in the grain boundary layer, $\varepsilon_0$ the permittivity in vacuum. Both $n_s$ and $\varepsilon_r$ vary with temperature. In $BaTiO_3$ based FE ceramics, $\varepsilon_r$ follows the Curie-Weiss law above $T_C$. Consequently, the rapid decrease of $\varepsilon_r$ above $T_C$ leads to a proportionally increased $\Phi$, thus an anomalous increase of resistivity $\rho$ ($\rho \propto \exp(\Phi/kT)$, creating the PTCR effect.

The width of depletion layer, b, is largest at about $T_C$, where all the surface acceptor states at grain boundaries are occupied. As temperature is increased above $T_C$, depopulation process of the occupied acceptor states begins to start as the acceptor energy is approaching the Fermi level. As a result, b decreases gradually as $n_s$ decreases. On the other hand, below $T_C$, the depletion layer will be refilled, more or less, by the charge compensation mechanism due to the incontinuous normal components of the spontaneous polarization at the grain boundary in the FE state. Therefore, b decreases rather sharply as temperature decreases below $T_C$. Such a variation of the width of depletion layer as a function of temperature in $BaTiO_3$ ceramics has been experimentally observed.[15]

In $BaTiO_3$ based high permittivity IBLC ceramics, below $T_C$, due to the charge compensation mechanism, the high capacitance depletion layer is more or less diminished, thus making ineffective, more or less, of an equivalent circuit of parallel capacitors. Consequently, the relative permittivity below $T_C$ is more governed by ferroelectric response, yielding a quite strong temperature dependent behavior. On the other hand, such an equivalent circuit will not be destroyed in the nonferroelectric perovskites CCTO and AFBO, where, the large relative permittivity persists with weak temperature dependence over a wide temperature range, until finally relax out at low temperature due to the MW relaxation which just originates from the effective serial-parallel circuit of the distinct grain and grain boundary barrier layer effects.[6]

The survival of the effective circuit in the FE state in our $Ba(Ti_{0.85}Zr_{0.15})O_3$:Ta ceramic samples, stems from its pseudo-cubic structure with very light rhombohedral distortion. The previous work in the literature concerned about the cubic-tetragonal FE phase transition in $BaTiO_3$ based ceramics. The appearance of large spontaneous polarization below cubic-tetragonal FE transition makes the charge compensation mechanism below $T_C$ more effective. On the other hand, as our x-ray data indicates, the rhombohedral distortion of our samples is very small, which means small ionic displacement from the centric position in the lattice. Hence, the cubic-rhombohedral FE transition in our samples is associated with a much lower spontaneous polarization value, which therefore is less effective to diminish the grain boundary barrier layers. Such a degradation of ferroelectricity in this doping level of $Ba(Ti,Zr)O_3$ rhombohedral phase can also manifest itself directly in the electric polarization behavior[12]. The observed phenomena in our study, i.e., the MW relaxation in the FE state, the weak temperature dependence (or, lack of feature) of the permittivity when passing $T_C$, are manifestation of the effective circuit in the FE state as in nonferroelectric CCTO and AFBO.

However, as our samples being ferroelectric, there still exist some features in the dielectric response which are different from the nonferroelectric counterparts. For example, the ε'(T) curves show comparatively more temperature dependence below $T_C$ and are more frequency dependent (Figure 2). With decreasing temperature, the magnitude of the high plateaus at low frequency side which is related to boundary layer property, doesn't remain nearly constant as observed in CCTO,[8-9] instead, shows a decrease, indicating the existence of charge compensation mechanism (Figure 3).

The low T mode relaxation arises from MW effect due to the effective circuit. The relaxation rate behavior (Figure 5) is found to be closely linked to the overall conductivity behavior which is shown in Figure 6. The strong upturn of the resistivity at around $T_C$ confirms the PTCR effect. Below $T_C$, it shows semiconductive behavior. At low temperatures, the resistance increases sharply and is beyond the measurable range below 120K. Below ~210 K, it shows a thermally activated conductivity process which follows the Arrhenius law $\rho=\rho_0\exp(\Delta_r/kT)$ with the activation energy $\Delta_r$=0.125ev. The value is close to the one ($\Delta_d$=0.128ev) for the low T mode relaxation at the same temperature range, indicating the close relationship between the low T mode relaxation and the conducting behavior in the grain. Above ~210 K, the resistivity behavior deviates from the Arrhenius law, which probably arises from the growing contribution from the grain boundary layer resistance. The deviation at ~210 K may be due to a critical change of the relative contribution of grain and grain boundary barrier layer. It is interesting to note that in the rhombohedral phase of Nb-doped $BaTiO_3$ single crystal, a slowing down of the relaxation motion at low temperatures which follows the Arrhenius law was also observed.[2]

Now let's turn to the high T mode relaxation (Figure 5). First we would note that in pure and doped $BaTiO_3$, a high frequency polar relaxation mode has been observed in the vicinity of $T_C$, with relaxation rate being minimum at $T_C$.[2,16] This was ascribed to the presence of order-disorder character in the usually accepted displacive picture for the FE phase transition in $BaTiO_3$. This type of local polar cluster (made up of off center $Ti^{4+}$ ions) induced relaxation mode occurs at very high frequencies (> $10^6$ Hz), and the relaxation rate above $T_C$ increases as temperature increases, which obviously is not responsible for the high T mode in our samples. Alternatively, we would relate it to the depletion layer in the FE ceramic samples.

The depletion layer at the grain boundary has been treated as a thin 2D layer. However, it should not always be true. As described above, the width of the depletion layer b is the largest at around $T_C$, and decrease gradually above $T_C$ resulting from the depopulation process of the occupied acceptor states at grain boundary, while decreasing sharply just below $T_C$ due to the charge compensation mechanism of FE phase transition. Experimental results have shown that in the vicinity of $T_C$, b is on the order of tenth of μm.[15] The typical grain size is on the order of μm. Taking into account of the fact that there are double depletion layers at both sides of a single grain boundary, it seems that in some cases, if not in all, the volume effect of the depletion layer area could not be ignored. The study of the dielectric property of the depletion layer in $BaTiO_3$ ceramics revealed that its equivalent dielectric constant (averaged value over the depletion layer) behaves similarly to that of the grain, namely a Curie-Weiss behavior above $T_C$ which is shifted 11 K downward relative to that of the grain.[17] From the temperature range in which the high T mode relaxation occurs, and the weaker and declining features of the magnitude, we would suggest that the high T mode arises from the depletion layer.

In our picture, we propose that the ionized oxygen vacancies do not distribute randomly in the depletion layer. The closer it is to the grain boundary, the less there are charged oxygen vacancies. This is very likely, due to two reasons. One, during the reoxidation along the grain boundary in the cooling process, some diffusion of the absorbed oxygen atoms from the grain boundary toward grain interior could happen. Two, the increasing height of the potential barrier when closer to the grain boundary, makes such a distribution of positively charged oxygen vacancies favorable in energy. It is thus expected that the distribution depends on the width of the depletion layer and the associated potential barrier. In such a picture, the slowing down of the relaxation rate of the high T mode with increasing temperature above $T_C$ can be understood as a result from the redistribution of the ionized oxygen vacancies in the depletion layer area when its width decreases with simultaneous change of the potential barrier height. Such a redistribution of the charges may enhance the local potential around the relaxing polar clusters which consist of off center $Ti^{4+}$ ions, due to Coulomb interaction. As a consequence, the relaxation motion slows down. And the gradual shrinkage of the depletion layer area above $T_C$ results in the declining of the amplitude of the high T mode relaxation.

## Conclusions

To sum up, we observed, in lightly doped n-type $BaTi_{0.85}Zr_{0.15}O_3$:Ta FE ceramics, a large relative permittivity spanning over a wide temperature range with weak temperature dependence, which finally relaxes out at low temperatures by a MW effect. The large dielectric response is caused by ceramic microstructures that consist of semiconductive grains and insulating grain boundaries, the same as in nonferroelectric perovskites CCTO, AFBO and ferroelectric $BaTiO_3$ based IBLC ceramics. However, the weaker temperature dependence and wider spanning temperature range, as well as the MW relaxation, differ from the conventional IBLC $BaTiO_3$ based ceramics. Moreover, another, and unusual, relaxation mode with slowed down relaxation rate as temperature rises, appears in the vicinity of $T_C$ and extends over a narrow temperature range. We suggest that these unusual phenomena can be understood as effects from the oxygen depletion layer at the grain boundary, in which the very slightly distorted rhombohedral crystal structure plays an important role. The unusual relaxation mode in the vicinity of $T_C$ is attributed to be from the volume effect of the depletion layer area which becomes unignorable when the PTCR effect takes place at FE phase transition, and the slowing down of the relaxation rate with increasing temperature is a consequence of redistribution of charged oxygen vacancies in the depletion layers.

**Acknowledgement** Part of this work was supported by the 21$^{st}$ century COE program 'Nanomaterial Frontier Cultivation for Industrial Collaboration' of the Ministry of Education, Culture, Sports, Science and Technology of Japan.

## References

(1) Samara, G. A. *J. Phys.: Condens. Matter* **2003**,*15,* R367

(2) Maglione, M.; Belkaoumi, M. *Phys. Rev. B* **1992**, *45*, 2029

(3) Bidault, O.; Maglione, M.; Actis, M.; Kchikech, M.; Salce, B. *Phys. Rev. B* **1995**, *52*, 4191

(4) Samara, G. A.; Boatner, L. A. *Phys. Rev. B* **2000**, *61*, 3889

(5) Chen Ang; Zhi Yu; Cross, L. E. *Phys. Rev. B* **2000**, *62*, 228

(6) Ion Bunget; Mihai Popescu, *Physics of Solid Dielectrics,* Elsevier, **1978**

(7) Subramanian, M. A.; Li, D.; Duan, N.; Reisner, B. A.; Sleight, A. W. *J. Solid State Chem.* **2000**, *151*, 323

(8) Sinclair, D. C.; Adams, T.; Morrison, F.; West, A. R. *Appl. Phys. Lett.* **2002**, *80*, 2153

(9) Homes, C. C.; Vogt, T.; Shapiro, S.; Wakimoto, S.; Ramirez, A. P. *Science*, **2001**, *293*, 673

(10) Raevski, I. P.; Prosandeev, S.; Bogatin, A.; Malitskaya, M.; Jastrabik, L. *J. Appl. Phys.* **2003**, *93*, 4130

(11) Cheng-Fu Yang. *Jpn. J. Appl. Phys*. **1996**, *35*, 1806

(12) Zhi Yu; Chen Ang; Ruyan Guo; Bhalla, A.S. *J. Appl. Phys*. **2002**, *92*, 1489

(13) Heywang, W. *Solid-State Electron*. **1961**, *3*, 51

(14) Jonker, G. H. *Solid-State Electron*. **1964**, *7*, 895

(15) Da Yu Wang; Kazumasa Umeya. *J. Am. Ceram. Soc*. **1990**, *73*, 669

(16) Maglione, M.; Boehmer, R.; Loidl, A.; Hoechli, U. T. *Phys. Rev. B* **1989**, *40*, 11441

(17) Da Yu Wang; Kazumasa Umeya. *J. Am. Ceram. Soc.* **1990**, *73*, 1574

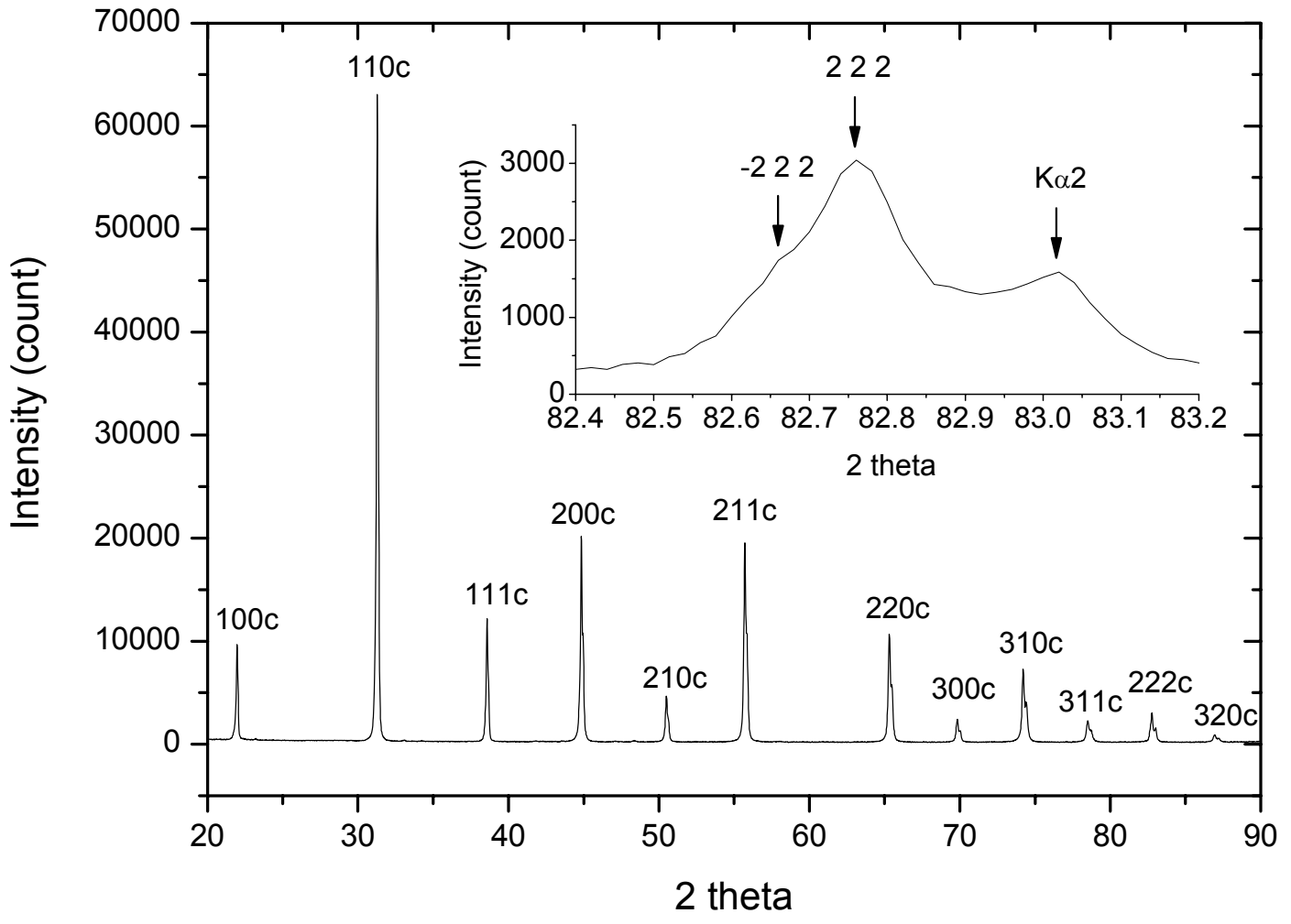

**Figure 1.** X-ray diffraction pattern recorded at room temperature for the sample BTZ:Ta008 (see text).

The peaks are indexed in a pseudo-cubic structure. Inset: a magnification around 222c peak.

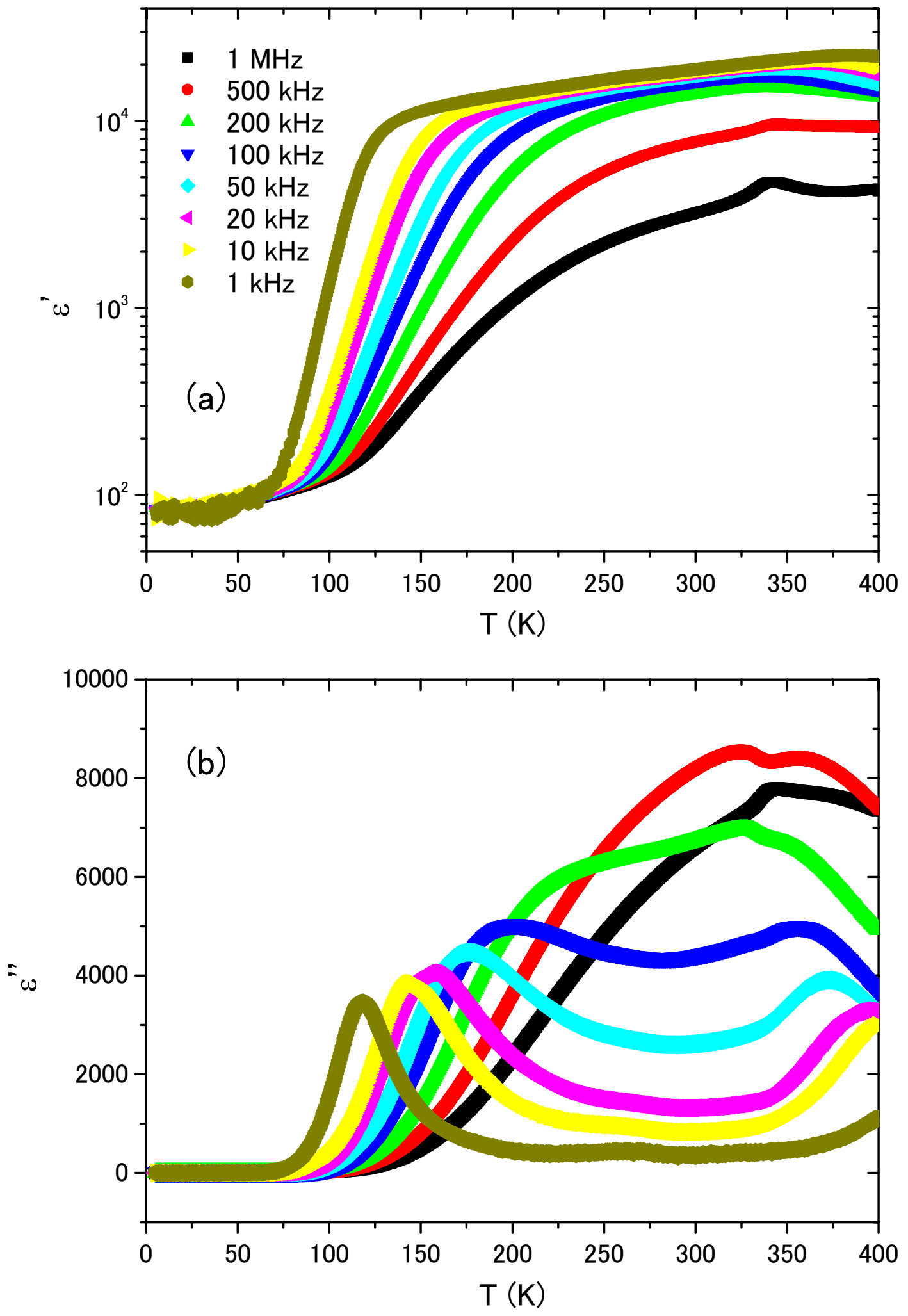

**Figure 2**. Temperature dependence of dielectric permittivity for BTZ:Ta008.

(a) real part ε', (b) imaginary part ε''.

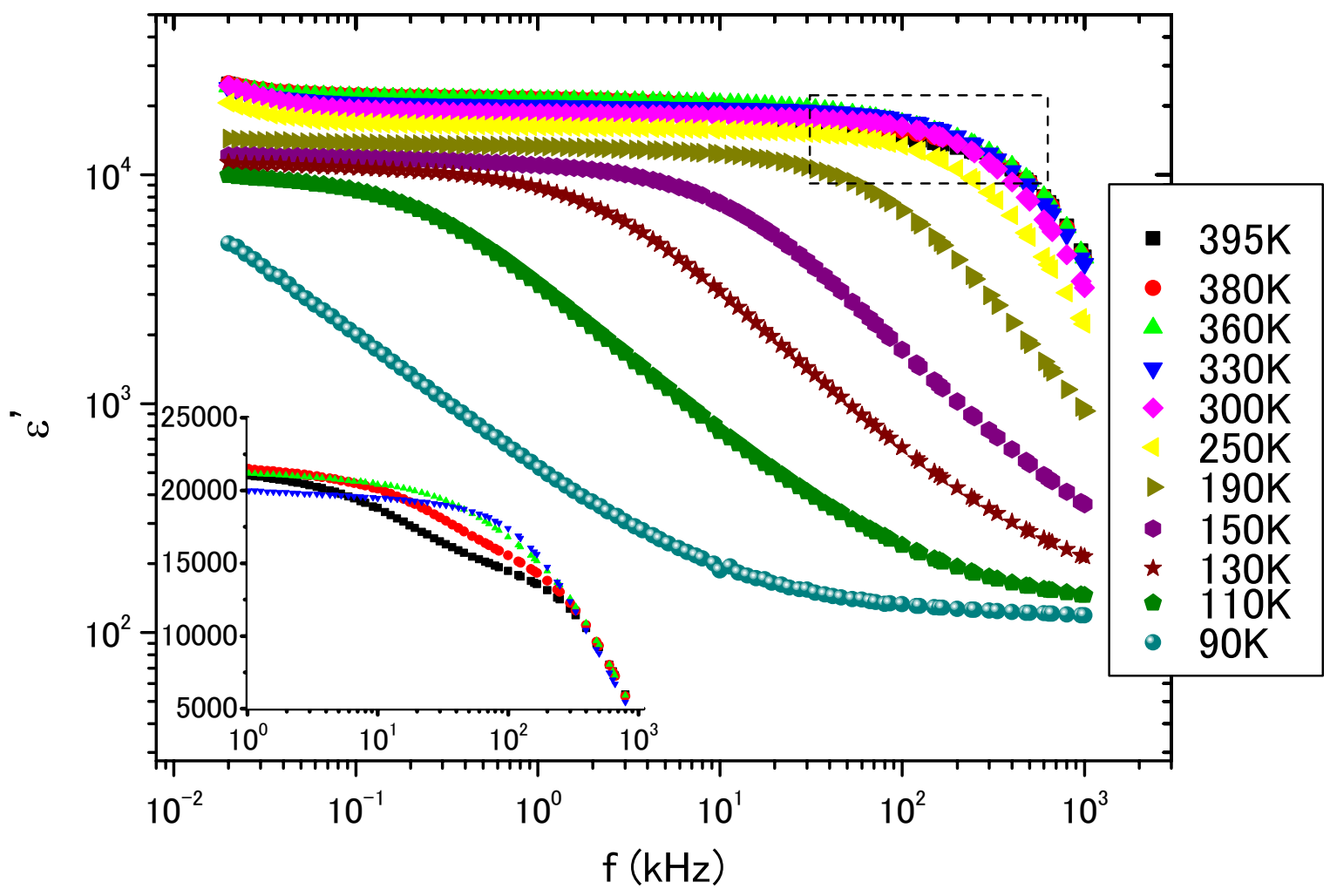

**Figure 3**. Frequency dependence of ε' at various temperatures. Inset: Expanded view of the marked area, highlighting the intermediate ε'(f) plateau for temperatures above 330K.

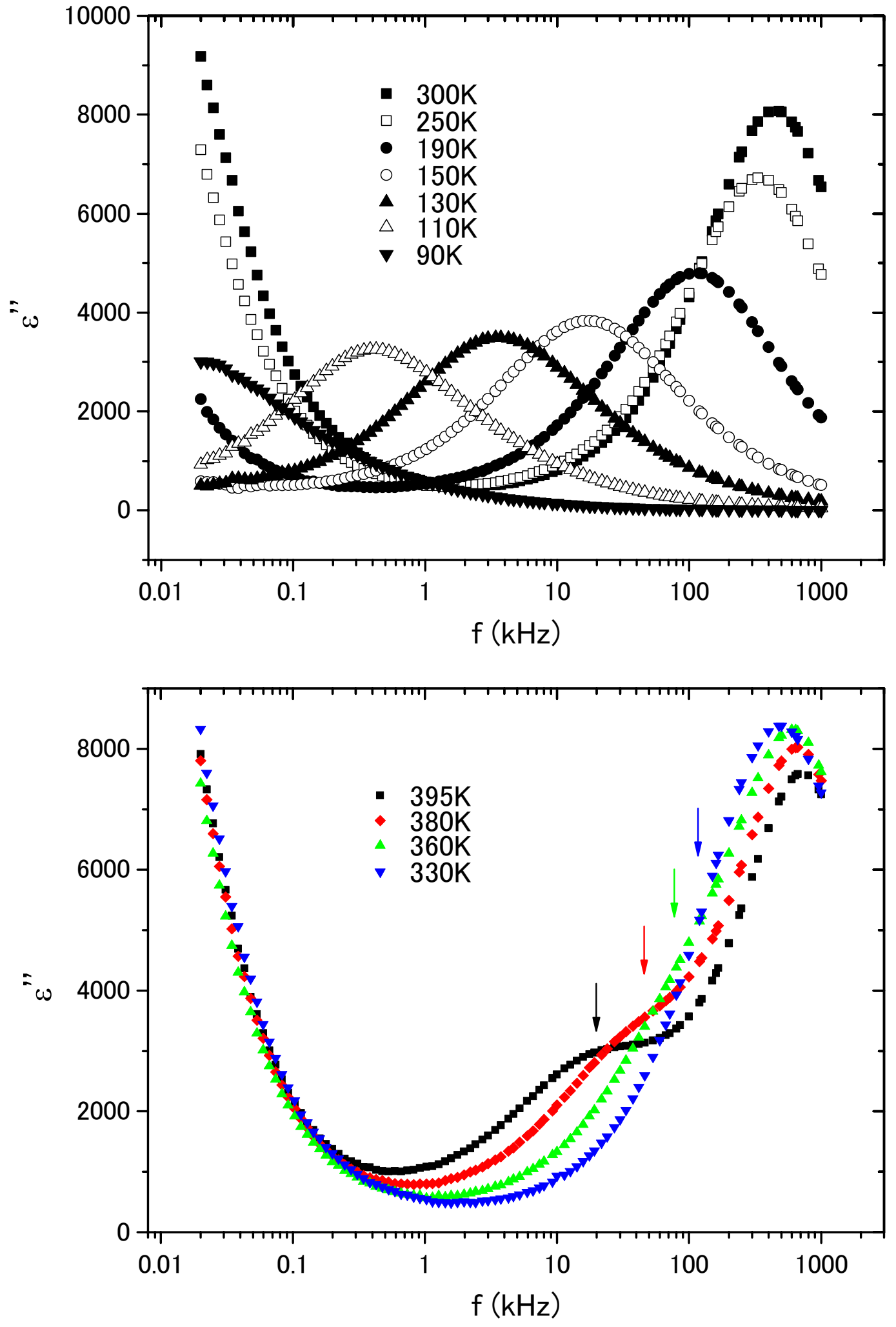

**Figure 4**. Frequency dependence of ε'' at various temperatures. (a) T<330K, (b) T≥330K. The arrows in (b) point to the weak peaks corresponding to the high T mode.

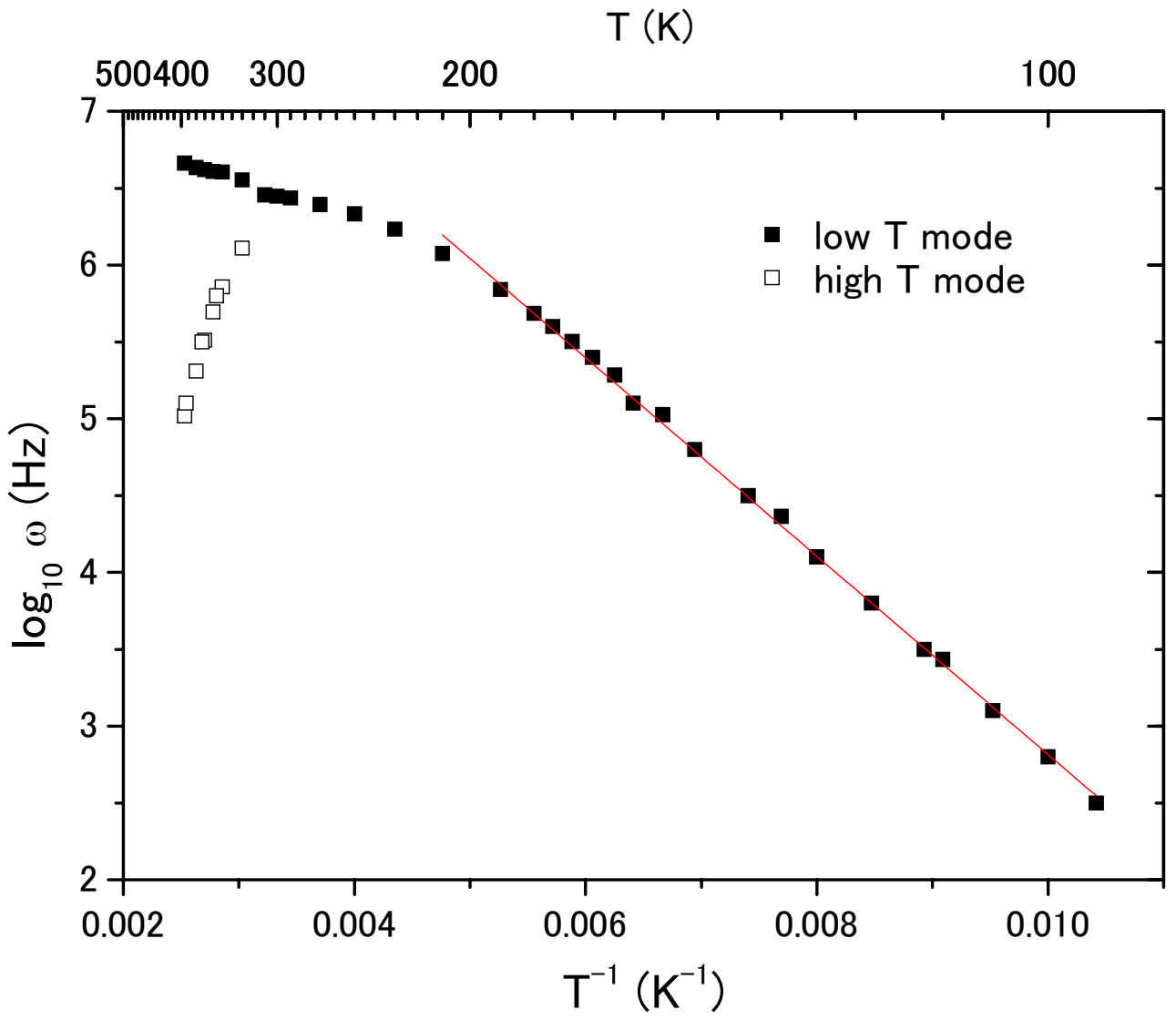

**Figure 5**. Arrhenius plot of relaxation rate ω versus $T^{-1}$.

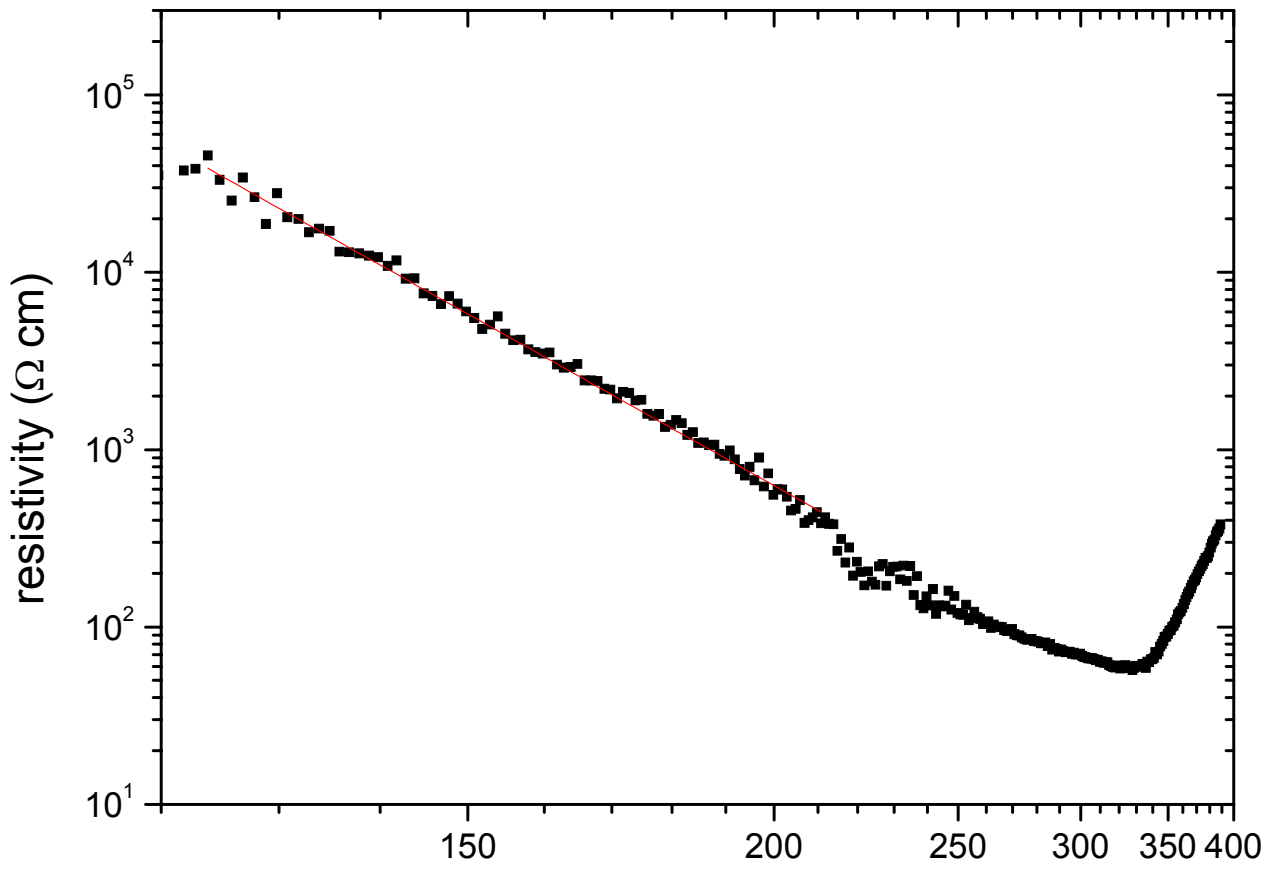

**Figure 6**. Temperature dependence of dc resistivity in a log-reciprocal plot.